\documentclass[letter,traditabstract]{aa} % for the long lists of affiliations 

\usepackage{graphicx}
\usepackage{txfonts}
\usepackage{natbib}
\begin{document}

\title{OH emission from  warm and dense gas in the Orion Bar PDR\thanks{{\em Herschel} is an ESA space
    observatory with science instruments provided by European-led
    Principal Investigator consortia and with important participation
    from NASA.}}

\author{Javier R. Goicoechea\inst{1},
Christine~Joblin\inst{2,3}, 
Alessandra~Contursi\inst{4}, 
Olivier~Bern\'e\inst{5}, 
Jos\'e~Cernicharo\inst{1},\\
Maryvonne~Gerin\inst{6}, 
Jacques~Le~Bourlot\inst{7}, 
Edwin~A.~Bergin\inst{8}, 
Tom A.~Bell\inst{1}
\and Markus~R\"ollig\inst{9}
}

\institute{Centro de Astrobiolog\'ia (CSIC/INTA), Ctra. de Torrej\'on a Ajalvir, km 4,
28850, Torrej\'on de Ardoz, Madrid, Spain. \\
\email{jr.goicoechea@cab.inta-csic.es}
%#2
\and 
Universit\'e de Toulouse, UPS-OMP, IRAP,  Toulouse, France.
%#3
\and 
CNRS, IRAP, 9 Av. colonel Roche, BP 44346, F-31028 Toulouse cedex 4, France
\and
Max-Planck-Institut f\"ur extraterrestrische Physik (MPE), Postfach 1312, 85741 Garching, Germany.
%#4
\and
Leiden Observatory, Leiden University, PO Box 9513, 2300 RA Leiden, The Netherlands. 
%#5
\and
LERMA, UMR 8112 du CNRS, Observatoire de Paris, \'Ecole Normale Sup\'erieure, France.
%#6
\and
Observatoire de Paris, LUTH and Universit\'e Denis Diderot, Place J. Janssen, 92190 Meudon, France.
%#7
\and Department of Astronomy, University of Michigan, 500 Church Street, Ann Arbor, MI 48109, USA.
%#8
\and  I. Physikalisches Institut, Universit\"at zu K\"oln,  Z\"ulpicher Str. 77, 50937 K\"oln, Germany.
%#9
}

\abstract{As part of a far-infrared (FIR)  spectral scan with {\em Herschel}/PACS, 
we present the first detection of the hydroxyl radical (OH) towards the Orion Bar photodissociation region (PDR).
 Five  OH (X $^2$$\Pi$; $\nu$=0) rotational $\Lambda$-doublets 
involving energy levels out to $E_u$/$k$$\sim$511\,K have been detected 
(at $\sim$65, $\sim$79, $\sim$84, $\sim$119 and $\sim$163\,$\mu$m).
The total intensity of the  OH lines is 
$\sum$$I$(OH)$\simeq$5$\times$10$^{-4}$~erg\,s$^{-1}$\,cm$^{-2}$\,sr$^{-1}$.
The observed emission of rotationally excited OH lines is extended and correlates well with 
the high-$J$ CO and CH$^+$~$J$=3-2 line emission (but apparently not with water vapour),
pointing towards a common  origin.
Nonlocal, non-LTE radiative transfer models 
including excitation
by the ambient FIR radiation field suggest that OH arises in a small filling factor
component of warm (T$_k$$\simeq$160--220\,K) and
dense  ($n_H$$\simeq$10$^{6-7}$\,cm$^{-3}$) gas with source-averaged OH column densities of $\gtrsim$10$^{15}$\,cm$^{-2}$. 
High density and temperature  photochemical models predict such enhanced OH columns at low  depths 
(A$_V$$\lesssim$1) and small spatial scales ($\sim$10$^{15}$\,cm),
where OH formation is driven by gas-phase endothermic reactions of atomic oxygen with molecular hydrogen.
We interpret the extended OH emission as coming from unresolved  structures exposed
to far-ultraviolet (FUV) radiation near the Bar edge
(photoevaporating clumps or filaments) and not from the lower density ``interclump'' medium. 
Photodissociation leads to OH/H$_2$O abundance ratios ($>$1) much higher 
than those expected in equally warm regions without enhanced FUV radiation fields.}

   \titlerunning{OH emission towards the Orion Bar PDR}
	\authorrunning{Goicoechea et al.}
   \maketitle
%
%________________________________________________________________

\section{Introduction}

In quiescent regions irradiated by cosmic-- or X--rays, the oxygen chemistry is initiated by the charge transfer from
H$^+$ and H$_{3}^{+}$ to atomic oxygen, forming O$^+$ and OH$^+$. 
In warm environments it can also start
with the reaction of atomic oxygen with H$_2$($\nu$=0) to form OH.
This  endothermic reaction (by $\sim$0.08\,eV) possesses an activation barrier 
of a few thousand~K and high gas temperatures ($\gtrsim$400\,K)
are needed to produce significant OH abundances (\textit{e.g.,}~in shocked gas).
In  molecular clouds exposed to strong far-ultraviolet (FUV) 
 radiation fields, the so--called PDRs,
the gas is heated to relatively high temperatures and there are also high
abundances of FUV-pumped vibrationally excited molecular hydrogen  H$_{2}^{*}$($\nu$$=$1,2...) \citep{hol97}. 
The internal energy available in H$_{2}^{*}$ 
can be used to overcome the O($^3P$)~+~H$_2$($\nu$=0) reaction  barrier \citep[see][and references therein]{agu10}.
Although not well constrained observationally, enhanced OH abundances are   expected in warm~PDRs. 

OH is a key intermediary molecule in the FUV-illuminated gas because further reaction of OH with 
H$_2$, C$^+$, O, N or S$^+$  leads to  the formation of H$_2$O, CO$^+$, O$_2$, NO or SO$^+$ respectively. 
Besides, OH is  the product of H$_2$O photodissociation,
the main destruction route of water vapour in the gas unshielded against FUV radiation.
Observations of OH in specific environments thus constrain different chemical routes of the oxygen chemistry.

Previous observations with  {\em KAO} and {\em ISO} 
have demonstrated that OH is a powerful tracer of the warm neutral gas in shocked gas;
from protostellar outflows and supernova remnants to extragalactic nuclei \citep[\textit{e.g.},][]{sto81,mel87,gon04}.
Unfortunately, the poor angular resolution ($>$1$'$) and sensitivity of the above telescopes  prevented us
from resolving the OH  emission from interstellar PDRs.

\begin{figure*}[t]
\vspace{-0.1cm}
\resizebox{\hsize}{!}{\includegraphics[angle=-90]{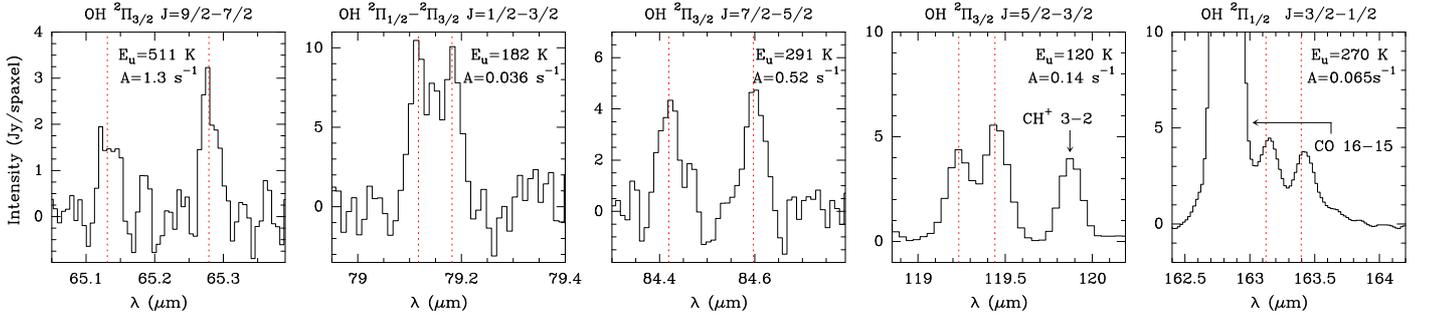}}
\caption{OH (X $^2$$\Pi$; $\nu$=0) rotational lines detected with \textit{Herschel}/PACS towards the 
$\alpha_{2000}$$\simeq$5$^h$35$^m$21.9$^s$, $\delta_{2000}$$\simeq$$-$5$^o$25$'$06.7$''$ position
where the higher excitation OH line peak is observed (see Figure~\ref{fig:OH_maps}).
The red dotted lines show the expected wavelength for each $\Lambda$-doublet
 (see Figure~\ref{fig_levels} in the appendix~\ref{sec:appenixB}  for a complete OH rotational energy diagram).
Small intensity asymmetries are observed in most OH $\Lambda$--doublets.
Transition upper level energies and $A_{ij}$ spontaneous transition probabilities
 are indicated.} 
\label{fig:OH_lines}
\end{figure*}

Owing to its proximity  
and nearly edge-on orientation, 
the interface region between the Orion Molecular Cloud~1 (OMC1) and the 
H\,{\sc ii} region illuminated by the Trapezium cluster, 
the  Orion Bar,
is the  prototypical warm PDR \citep[with a FUV radiation field
 at the ionization front of $\chi$$\simeq$2.5$\times$10$^4$ times the mean interstellar 
field in Draine units;][]{mar98}.
The most commonly accepted scenario is that 
an extended gas component, with mean gas densities $n_H$ of 10$^{4-5}$\,cm$^{-3}$,
causes the chemical stratification seen in the PDR \citep{hog95}. 
Most of the low-$J$ molecular  line emission arises in this extended ``interclump'' medium \citep{tie85,sim97,wie09,hab10}.
In addition, another component  of  higher density clumps
 was introduced to fit
the observed  H$_{2}$, high--$J$~CO, CO$^+$ and other high density and temperature  tracers 
\citep{bur90,par91,sto95,wer96,you00}.
Owing their small filling factor this clumpy structure would allow FUV radiation to permeate the region.
The presence of dense clumps is  still controversial. 

In this letter we present initial results from a  spectral scan of the Orion Bar
taken with  the PACS instrument \citep{pog10} on board $\textit{Herschel}$ \citep{pil10} 
as part of  the ``HEXOS'' key programme \citep{ber10}. 
PACS observations of OH lines towards young stellar objects have recently been reported
by Wampfler et al. \citep{wam10}.
Here we present the first detection of OH towards this
prototypical PDR.

\vspace{-0.2cm}
\section{Observations and data reduction}

PACS observations were carried out on 7 and 8 September 2010
and consist of two spectral scans in Nyquist sampling wavelength
range spectroscopy mode.
The PACS spectrometer uses photoconductor detectors and  provides 25 spectra  over a 47$''$$\times$47$''$ field-of-view
resolved in 5$\times$5 spatial pixels (``spaxels''), each with a size of $\sim$9.4$''$$\times$9.4$''$ on the sky.
The measured width of the spectrometer point-spread function (PSF) is relatively constant at
$\lambda$$\lesssim$125\,$\mu$m but it  increases above the spaxel size for longer wavelengths. 
The resolving power varies between $\lambda$/$\Delta$$\lambda$$\sim$1000 (R1 grating order) and $\sim$5000  
(B3A). 
The central spaxel was centred at  
$\alpha_{2000}$: 5$^h$35$^m$20.61$^s$, $\delta_{2000}$: -5$^o$25$'$14.0$''$ target  position.
Observations were carried out in the  ``chop-nodded'' mode with the largest chopper throw of 6~arcmin.
Nominal capacitances (0.14\,pF) were used.
The integration time was 3.2\,h for the 1342204117 observation (B2B and R1) 
 and 2.7\,h for the 1342204118 observation (B3A and R1). 
The data were processed with HIPE using a pipeline upgraded with a spectral flatfield
algorithm that reduces the spectral fringing seen in the  Nyquist-sampled wavelength spectra of bright sources.
Figure~\ref{fig:OH_lines} shows the resulting OH lines towards the OH emission peak and Figure~\ref{fig_correlations} 
shows the 
intensities measured in each of the 25 spaxels for several lines of OH, CO, CH$^+$, H$_2$O and [N\,{\sc{ii}}]. 
In order to better sample the PSF and obtain accurate line intensities to be reproduced with our models,  
we fit the  OH  emission  averaged over several adjacent spaxels in 
Section~\ref{sec:excitation} (see also appendix~\ref{sec:appenixA}).

\begin{figure}[h]
\vspace{-0.2cm}
\includegraphics[width=8.cm, angle=-90]{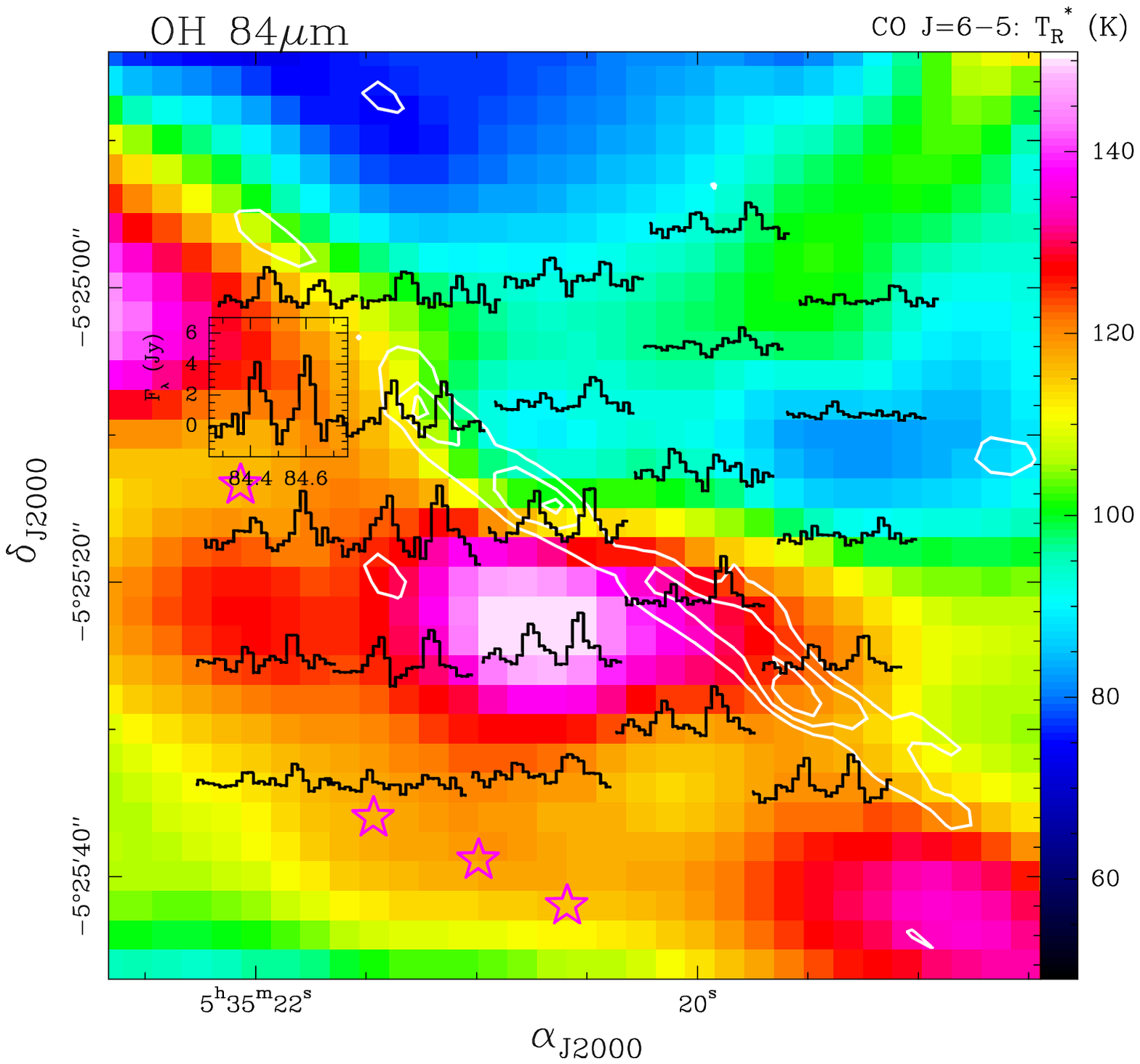}
\caption{PACS rotationally excited
 OH  $^{2}\Pi_{3/2}$ $J$=7/2$\rightarrow$5/2 lines at $\sim$84\,$\mu$m ($E_u$/$k$=291\,K)
overlaid on the
distribution of the CO $J$=6-5 peak brightness temperature (colour image)
observed with the CSO telescope at $\sim$11$''$  resolution \citep{lis98}.
White contours show the brightest regions of H$_{2}^{*}$ $v$=1-0 $S$(1) emission \citep{wal00}. 
Lower intensity H$_{2}^{*}$ extended emission is
present in the entire field \citep{wer96}.
Violet stars shows the position of the H$^{13}$CN $J$=1-0  clumps deeper inside the Bar
\citep{lis03}. Note the decrease of OH line intensity
with distance from the ionization front.}
\label{fig:OH_maps}
\end{figure}

\vspace{-0.2cm}
\section{Results}

Of all the observed OH lines only the ground-state lines at $\sim$119\,$\mu$m  show widespread bright emission 
at all positions. The $\sim$119\,$\mu$m lines mainly arise from the background OMC1 complex
(the same applies to most ground-state lines of other species).
Figure~\ref{fig:OH_maps} shows the spatial distribution of the
rotationally excited OH $^{2}\Pi_{3/2}$ $J$=7/2$\rightarrow$5/2 lines at $\sim$84\,$\mu$m ($E_u$/$k$=291\,K)
superimposed over the CO $J$=6-5 peak brightness temperature \citep[colour image from][]{lis98} 
and over the brightest  H$_{2}^{*}$ $\nu$=1-0 $S$(1) line emission regions
\citep[white contours from][]{wal00}.
The emission from the other  OH $\Lambda$-doublets at $\sim$79 and $\sim$163\,$\mu$m 
(see Figure~\ref{fig:OH_maps2} in appendix~\ref{sec:appenixB}) 
displays a similar spatial distribution that follows the ``bar'' morphology   peaking
near the H$_{2}^{*}$~$v$=1-0 $S$(1) bright emission region and then decreases with distance from the
ionization front \citep[note that H$_{2}^{*}$ shows lower level extended
emission with small-scale structure in the entire observed field;][]{wer96}.
The excited OH spatial distribution, however,  does not follow the CO~$J$=6-5 emission maxima, 
which approximately trace the gas temperature 
in the extended ``interclump'' component.

Figure~\ref{fig:OH_lines} shows  the detected OH $\Lambda$-doublets 
(at $\sim$65, $\sim$79, $\sim$84, $\sim$119 and $\sim$163\,$\mu$m) towards the 
position where the higher excitation OH lines peak. 
The total intensity of the observed FIR lines 
 is $\sum$$I$(OH)$\simeq$5$\times$10$^{-4}$~erg\,s$^{-1}$\,cm$^{-2}$\,sr$^{-1}$.
 All OH doublets appear in emission, with intensity asymmetry ratios up to 40\% 
(one line of the $\Lambda$-doublet is  brighter than the other).

Note that the upper energy level of the $^{2}\Pi_{3/2}$~$J$=9/2$\rightarrow$7/2 transition 
at $\sim$65\,$\mu$m lies at  E$_u$/$k$$\sim$511\,K.
The critical densities ($n_{cr}$) of  the observed  OH transitions are high,
$n_{cr}$$\gtrsim$10$^{8}$\,cm$^{-3}$. For much lower gas densities, and in the presence of strong
FIR radiation fields,
most lines would have been observed in absorption, especially those in the $^{2}\Pi_{3/2}$ ladder \citep{goi02}.
Hence, the observed OH lines  must
arise in an widespread component of warm and dense gas.

\begin{figure} [t]
\vspace{-0.0cm}
\resizebox{\hsize}{!}{\includegraphics[angle=-90]{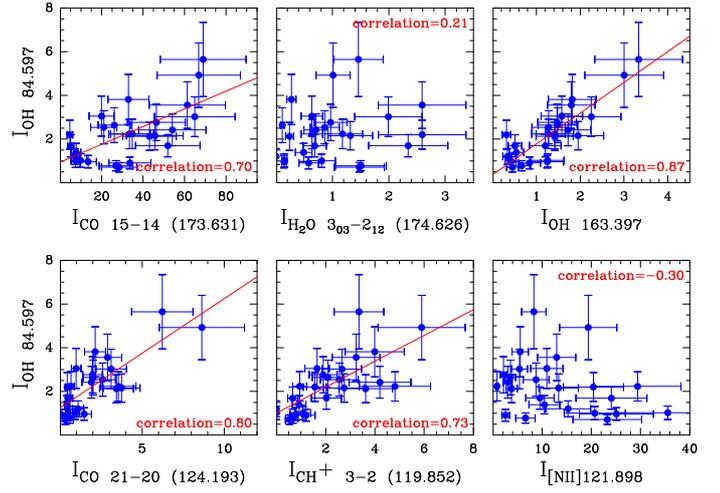}}
\caption{Line intensity spatial correlations between the  OH 
$^{2}\Pi_{3/2}$ $J$=7/2$^-$$\rightarrow$5/2$^+$  line at 84.597\,$\mu$m 
and lines from other species.
Intensities are in units of 10$^{-5}$\,erg\,cm$^{-2}$\,s$^{-1}$\,sr$^{-1}$.}
\label{fig_correlations}
\end{figure}

Although our PACS observations do not provide a fully sampled map,
 the line emission observed in the 25 spaxels 
can be used to carry out a first-order analysis on the spatial correlation of  different
 lines (neglecting perfect PSF sampling, line opacity and excitation effects).
Except for the OH ground-state lines at $\sim$119\,$\mu$m (that come from the background OMC1 cloud),
we find that the rotationally excited OH lines correlate well with the high-$J$ CO and CH$^+$ emission
but, as expected, they do not correlate with the ionized gas emission.
Figure~\ref{fig_correlations} (\textit{lower panel}) compares the observed OH~$\sim$84.597\,$\mu$m
line intensities with those of 
CO $J$=21-20 ($E_u$/$k$$\sim$1276\,K), CH$^+$ $J$=3-2 ($E_u$/$k$$\sim$240\,K)
and [N\,{\sc{ii}}]121.891\,$\mu$m  (all observed with a similar PSF)
and also with the  CO $J$=15-14 ($E_u$/$k$$\sim$663\,K), H$_2$O 3$_{03}$-2$_{12}$ ($E_u$/$k$$\sim$163\,K) and
OH~$\sim$163.397\,$\mu$m lines (\textit{upper panel}).
This simple analysis suggests
that the excited OH, high-$J$ CO and CH$^+$ $J$=3-2 lines  arise from the same gas  component.
It also shows that the emission from different excited OH lines is well correlated,
while the OH and H$_2$O  emission is not (within the PSF sampling caveats).

\vspace{-0.2cm}
\section{OH column density determination}
\label{sec:excitation}

Determining the OH level populations is no  trivial excitation problem.
In addition to relatively strong asymmetries in the collisional 
rate coefficients\footnote{We used  
collision rate coefficients of OH with $para$- and $ortho$-H$_2$  
from  Offer \& van Dishoeck \citep{off92} and Offer et al. \citep{off94}.
Strong differences in the intensity of
each OH $\Lambda$-doublet component due to  asymmetries in the collisional
rates between OH and $para$-H$_2$ were predicted
 (\textit{e.g.,}~$I$(84.597)/$I$(84.420)$>$$I$(119.441)/$I$(119.234)$>$1).
Asymmetries are significantly reduced when collisions with $ortho$-H$_2$ are included
 (\textit{i.e.,}~in the warm gas) and when FIR radiative pumping plays a role.
We assume that the H$_2$ $ortho$-to-$para$ ratio is thermalized to the gas temperature, 
\textit{e.g.,}~$\sim$1.6 at 100\,K and $\sim$2.9 at 200\,K.} between
each $\Lambda$-doubling component (Offer \& Van Dishoeck 1992),
radiative and opacity effects (pumping by the ambient IR radiation field and line trapping) can play a significant role
if the  gas density is much lower than~$n_{cr}$.
Here we use a nonlocal
and non-LTE code that treats both the OH line 
and continuum radiative transfer \citep[see appendix in][]{goi06}.
The continuum measured by PACS and SPIRE in the region (H. Arab et al. in prep.)
 can be approximately reproduced by a modified blackbody with a colour temperature of $\sim$55\,K and a
opacity dependence of $\sim$0.05(100/$\lambda$)$^{1.75}$ \citep[see][]{lis98}.
Our calculations include thermal, turbulent, and
opacity line broadening with a turbulent velocity dispersion of
$\sigma$=1.0\,km\,s$^{-1}$ and FWHM=2.355$\times\sigma$
\citep[see \textit{e.g.}, the  linewidths measured by][]{hog95}. 
A grid of single-component models for different $N$(OH), gas temperatures, densities and beam filling factors  
were run. 
The best fit model was found by minimizing the ``$\chi^2$-value" (see the appendix~\ref{sec:appenixA} for its definition).

From the excitation models we conclude that the
high $I$(65.279)/$I$(84.596)$\simeq$0.3 and $I$(65.132)/$I$(84.420)$\simeq$0.5  intensity ratios in the $^{2}\Pi_{3/2}$ ladder
can only be reproduced if the gas is dense, at least $n_H$=$n$(H)+2$n$(H$_2$) of a few 10$^6$\,cm$^{-3}$.
In addition, no model is able to produce even a crude fit to the data if one assumes 
the average molecular gas temperature  (T$_k$$\simeq$85\,K)  in the lower density ``interclump'' medium \citep{hog95}.
In the $^{2}\Pi_{1/2}$ ladder, the intensity of the $\sim$163\,$\mu$m lines is sensitive to the 
FIR radiation field in the region through the absorption of FIR dust continuum photons in the
OH\,$\sim$34 and $\sim$53\,$\mu$m cross-ladder transitions ($^{2}\Pi_{3/2}$-$^{2}\Pi_{1/2}$ $J$=3/2$\rightarrow$5/2
and $J$=3/2$\rightarrow$3/2 respectively).   
However, the $\sim$163\,$\mu$m lines in the Orion Bar are not particularly strong and the OH~$\sim$34 and $\sim$53\,$\mu$m lines
are not present in the ISO spectra,  
thus FIR pumping does not dominate the OH excitation. 
All in all, the best model is found for a source of high density ($n_H$$\lesssim$10$^7$~cm$^{-3}$)
and warm gas temperatures (T$_{k}$=160-220\,K).  
This temperature lies in between
the  $\sim$600\,K derived in the H$_2$ emitting regions \citep{all05}
and the $\sim$150\,K derived from  NH$_3$ lines \citep{bat03}  
 near the ridge of dense and cooler H$^{13}$CN clumps \citep[T$_k$$\simeq$50\,K;][]{lis03}.  
In our simple model, the best fit is obtained for a source of small filling factor ($\eta\simeq$10$\%$)
with a source-averaged OH column density of $\gtrsim$10$^{15}$\,cm$^{-2}$ (see~appendix~\ref{sec:appenixA}).

\vspace{-0.2cm}
\section{OH chemistry and line emission origin}
\label{sec:PDRmods}

\begin{figure}[t]
\vspace{-0.00cm}
\resizebox{\hsize}{!}{\includegraphics[angle=-90]{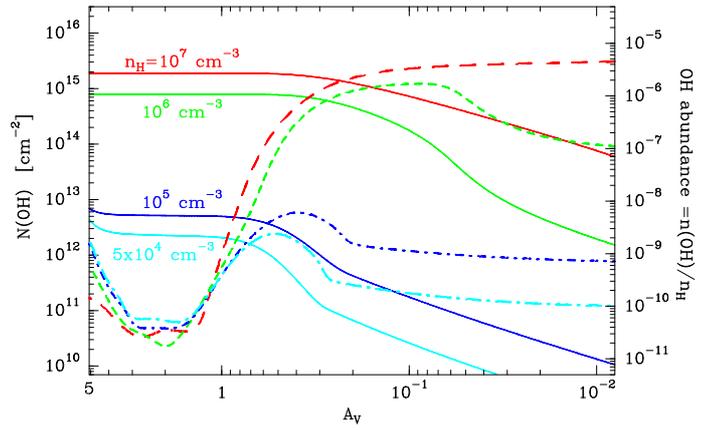}}
\caption{Gas-phase PDR models of a FUV-illuminated slab ($\chi$=10$^4$) of gas with
 different densities:
$n_H$=10$^7$, 10$^6$, 10$^5$ and 5$\times$10$^4$\,cm$^{-3}$. 
OH column densities are shown as continuous curves (left axis) 
while OH abundances, $n$(OH)/$n_H$, are shown as dashed curves (right axes).}
\label{fig:pdr_mod}
\end{figure}

We  used the Meudon PDR code \citep{lpt06,goi07} to estimate the OH column 
density in a slab of gas at  different densities ($n_H$ from 5$\times$10$^4$ to 10$^7$\,cm$^{-3}$). 
The adopted FUV radiation field, $\chi$=10$^4$, roughly corresponds to the attenuation
of the FUV field at the ionization front  by a column density of $N_H$$\simeq$10$^{21}$\,cm$^{-2}$ in
a $n_H$$\simeq$10$^4$\,cm$^{-3}$  medium. This attenuation is equivalent to a spatial length
of $\sim$10$^{17}$\,cm, consistent with the observed  decrease of excited
 OH emission with projected distance from the ionization front (see Figure~\ref{fig:OH_maps}).
Given the high gas temperature, FUV field and moderate grain temperature 
(T$_{gr}$$\sim$70-100\,K) in the regions traced by  
FIR OH, CO and CH$^+$ lines, we neglected molecule freeze-out and ice desorption
\citep[which are important deeper inside at A$_V$$\gtrsim$3;][]{hol09}.
In our models   OH is a surface tracer that reaches its  peak abundance at A$_V$$\lesssim$1
(Figure~\ref{fig:pdr_mod}) where OH formation is driven by the endothermic reaction O($^3P$)~+~H$_2$~$\rightarrow$~OH~+~H, 
slightly enhanced by the O($^3P$)~+~H$_{2}^{*}$ reaction 
\citep[included in our models; see][]{agu10}.
Gas temperatures around $\sim$1000-500\,K are predicted near the slab surface at A$_V$=0.01
and around $\sim$100\,K at A$_V$$\gtrsim$1. In these H/H$_2$ transition layers where the OH abundance peaks,
the electron density is still high ($\lesssim$[C$^+$/H]\,$n_H$) and hydrogen is not  fully molecular, 
with  $f$(H$_2$)=2$n$(H$_2$)/[$n$(H)+2$n$(H$_2$)]$\simeq$0.5.
In general, the higher the gas temperature  where enough H$_2$ has formed, 
the higher  the predicted OH abundance.

In the A$_V$$\lesssim$1 layers,  OH destruction is dominated by photodissociation 
(OH~+~$h\nu$~$\rightarrow$~O~+~H)
and to a  lesser extent, by reactions of OH with H$_2$ to form H$_2$O 
(only when the gas temperature and density are very high).
Water vapour photodissociation (H$_2$O~+~$h\nu$~$\rightarrow$~OH~+~H) in the surface layers limits the H$_2$O formation
  and leads to OH/H$_2$O abundance ratios ($>$1), much higher than those expected in equally warm regions without 
enhanced FUV radiation fields (\textit{e.g.} in C--shocks).
The lack of apparent correlation between the excited OH and H$_2$O 3$_{03}$-2$_{12}$ lines 
(see Figure~\ref{fig_correlations}) and the
absence of high excitation H$_2$O lines in the PACS spectra (only weak H$_2$O 2$_{21}$-2$_{12}$, 2$_{12}$-1$_{01}$ 
and 3$_{03}$-2$_{12}$ lines are clearly detected) suggests
that the bulk of OH and H$_2$O column densities arise from different  depths.

As the temperature decreases inwards, the gas-phase production of OH also decreases.
The  spatial correlation between excited OH, CH$^+$ $J$=3-2  and high-$J$ CO lines is a good 
signature of their common origin in the warm gas at low $A_V$ depths.

Our PDR models predict OH column densities in the range  $\sim$10$^{12}$\,cm$^{-2}$ to $\sim$10$^{15}$\,cm$^{-2}$
at  A$_V$$\lesssim$1
for gas densities between $n_H$=5$\times$10$^4$ and 10$^7$\,cm$^{-3}$ respectively.
Hence, even if we take into account possible inclination 
effects, high density and temperature models 
produce OH columns closer to the values derived in Section~4 just from excitation
considerations (note that a precise determination of the
gas density would require knowing the collisional rate coefficients of OH with H atoms and electrons). 
The OH abundance in these dense surface layers is of the order of $\approx$10$^{-6}$ with respect to H nuclei.
However, optical depths of A$_V$$\lesssim$1 correspond to spatial scales of only $\sim$10$^{15}$\,cm
\citep[\textit{i.e.}, much smaller than the H$^{13}$CN clumps detected by][deeper
inside the cloud]{lis03},
but we detect extended OH emission over $\sim$10$^{17}$\,cm scales. Therefore, we  have to conclude that 
the observed OH emission arises from a small filling factor ensemble of unresolved structures 
that are exposed to FUV radiation (overdense clumps or filaments). 
Note that owing to the lower grain temperature compared to the gas, 
the expected FIR continuum emission from these clumps will still be below the continuum
levels observed by \textit{Herschel}/PACS.

The minimum size of the dense clumps is $\sim$10$^{15}$\,cm (from OH photochemical models)
with a maximun size of $\sim$10$^{16}$\,cm (from the inferred beam dilution factor). Both correspond
to $\lesssim$0.2$''$-2$''$ at the distance of Orion.
As an example, H$_2$ photoevaporating clumps of $\sim$10$^{16}$\,cm size have been unambiguously resolved 
towards  S106 PDR \cite{noe05}. However, higher angular resolution observations 
(\textit{e.g.,}~with 8-10m  telescopes) are needed
to resolve smaller H$_2$ clumps from the H$_2$ interclump emission in the Orion Bar.

If the observed FIR OH line emission does not arise from such a high density gas component, a different non-thermal 
excitation mechanism able to populate the  OH $^{2}\Pi_{3/2}$ $J$=9/2 and 7/2 levels would be needed.
Two alternative scenarios can be explored, at least qualitatively.
First, OH molecules produced by H$_2$O photodissociation are expected to form mostly in the ground electronic
and vibrational state but in unusually high energy $J$$>$70/2 levels \citep[a few thousands~K!;][]{van00}. 
Nevertheless, they   
 will cascade down radiatively to  lower energy rotational ladders extremely rapidly.
Although very excited suprathermal OH $J$$\simeq$70/2-15/2 lines have been reported
towards the HH\,211 outflow
and were interpreted as H$_2$O photodissociation \citep{tap08},  we
did not find any of them in the SWS or in the IRS spectra of the Orion Bar
(taken and processed by us from the ISO and \textit{Spitzer} basic calibrated data archives). 
Besides, even if photodissociation is the main H$_2$O destruction mechanism in the A$_V$$\lesssim$1 warm layers, 
this  is not the main production pathway of OH (the O($^3P$)~+~H$_2$ reaction dominates).

Second,  experiments and quantum calculations  suggest that reaction of O($^3P$) atoms with H$_{2}^{*}$($\nu$=1)  
can produce significant OH in the $\nu$=1 vibrationally excited state, with OH($\nu$=1)/OH($\nu$=0) 
population ratios $\gtrsim$1 for moderate kinetic energies
\citep[$\gtrsim$0.05-0.1\,eV; \textit{e.g.,}][]{bal04}.
Complementarily, absorption of near-- and mid--IR photons (from the continuum or from bright overlapping H$_{2}^{*}$ 
and ionic lines) can also pump OH to the $\nu$=1 state.
In both cases, subsequent de-excitation  through the $\nu$=1-0 rotation--vibration band 
at $\sim$2.80\,$\mu$m
would  populate the  OH($\nu$=0) rotationally excited levels  that we observe with PACS.
However, the OH $\nu$=1-0 band at $\sim$2.80\,$\mu$m is not present in the ISO/SWS spectra.
Even assuming that the O($^3P$)+H$_{2}^{*}$($\nu$=1) reaction only forms OH($\nu$=1),
significant gas column densities at very hot temperatures (T$_k$$\sim$2000\,K) will be needed to match
the observations if  $n_H$$\simeq$10$^{4-5}$\,cm$^{-3}$.
Studying the OH vibrationally-pumping mechanism quantitatively is beyond the scope of this work 
but these chemical and pumping  effects could contribute to the excitation of  FIR OH lines
in lower density gas.

Different scenarios for the origin and nature of photoevaporating clumps have been proposed
\citep{gor02}, but without a more precise determination of their
sizes and densities it is hard to conclude on any of them. Subarcsec resolution observations of OH gas-phase products, 
\textit{e.g.,}~direct observation of CO$^+$ or SO$^+$ clumps with \textit{ALMA},
will help us to assay the clumpy nature of the Orion Bar in the near future.

\begin{acknowledgements}
PACS has been developed by
a consortium of institutes led by MPE (Germany)
and including UVIE (Austria); KU Leuven, CSL,
IMEC (Belgium); CEA, LAM (France); MPIA (Germany); 
INAF-IFSI/OAA/OAP/OAT, LENS, SISSA (Italy); IAC (Spain). 
We thank Darek Lis and Malcolm Walmsley for providing us their CO $J$=6-5 and H$_{2}^{*}$~1-0~S(1) maps.
We also thank Emilie Habart, Bart Vandenbussch and Pierre Royer  for useful discussions and  the referee 
for his/her constructive report.
We thank the Spanish MICINN for funding support
through grants AYA2006-14876, AYA2009-07304 and CSD2009-00038. 
JRG is supported by a Ram\'on y Cajal research contract.
 We acknowledge the use of the LAMDA data base \citep{sch05}. 

\end{acknowledgements}

\clearpage

\begin{appendix}

\section{$\chi^2$ analysis and intensity extraction}
\label{sec:appenixA}

\begin{figure}[ht]
\begin{center}
\resizebox{\hsize}{!}{\includegraphics[angle=-90]{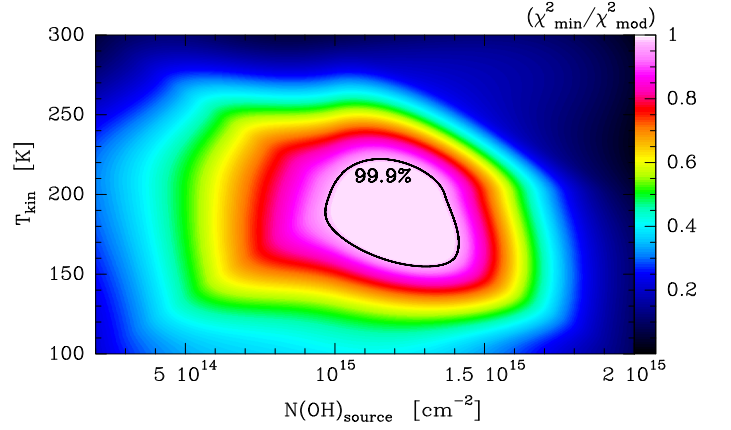}}
\end{center}
\caption{$\chi^{2}_{min}$/$\chi^{2}_{mod}$ as a function of T$_{k}$
and $N$(OH) for a grid of models with 
$n_H$=10$^7$\,cm$^{-3}$ and  $\eta$=0.1.
The set of parameters 
giving the minimum $\chi^2$-values with a confidence level of 99.9$\%$  are shown.}
\label{fig_x2}
\end{figure}

The best-fit OH radiative transfer model was obtained
by finding the minimum  $\chi^2$--value defined as
\begin{equation}
\chi^2 = \frac{1}{n-p} \sum_{i=1}^{n} \left(\frac{I_{obs}^{i}-I_{mod}^{i}}{\sigma_{obs}^{i}}\right)^2, 
\label{eq-tot-diag}
\end{equation}   
where $n$ is the number of observed OH lines, $p$ is the number of free parameters in the fit,
$I_{obs}^{i}$ and $I_{mod}^{i}$  are the observed and modelled line integrated intensities,
and $\sigma_{obs}^{i}$ is  the 1$\sigma$ uncertainty of $I_{obs}^{i}$. Therefore, we fit
absolute intensity values and not line intensity ratios.
Figure~\ref{fig_x2} shows the best grid of radiative transfer models with a density
of $n_H$=10$^7$\,cm$^{-3}$ and a beam filling factor $\eta$=0.1,
defined as $\eta = \frac{\Omega_{s}}{\Omega_{s}+\Omega_{PSF}}$, 
where $\Omega_{s}$ and $\Omega_{PSF}$ are the sky solid angles subtended by the source
and by the PACS PSF respectively. 
Figure~\ref{fig_x2} shows the best source-averaged OH column densities,
where $N$(OH)$_{beam}$$\approx$\,$\eta\,\times$\,$N$(OH)$_{source}$.

The list of OH line intensities around the 
position where the higher excitation OH lines peak 
is shown in Table~\ref{table:intens}. 
In order to better sample the spectrometer PSF in these extended emission observations, 
the OH line intensities were
 computed by co-adding the measured fluxes  over four adjacent spaxels
(the [3,4], [3,3], [2,4] and [2,3]). Owing to the broader PSF width
at long wavelengths, the entire array was used
to extract the OH~$\sim$163\,$\mu$m line intensities.

 \begin{table}[h]
      \caption[]{PACS OH line intensities towards the Orion Bar PDR$^a$.}
\centering
\label{tab1}
\begin{tabular}{ccc}
\hline\hline
\multicolumn{1}{c}{$\lambda$} & 
\multicolumn{1}{c}{OH} &
\multicolumn{1}{c}{$I_{obs}$ ($\sigma_{obs}$)}  \\
 ($\mu$m) & transition & ($\times$10$^{-5}$ erg\,s$^{-1}$cm$^{-2}$sr$^{-1}$) \\
\hline\hline
         & &    \\
119.441  & $^{2}\Pi_{3/2}$ $J$=5/2$^+$-3/2$^-$ &10.09 (0.21)\\
119.234  & $^{2}\Pi_{3/2}$ $J$=5/2$^-$-3/2$^+$ & 9.44 (0.21)\\
 84.597  & $^{2}\Pi_{3/2}$ $J$=7/2$^-$-5/2$^+$ & 4.07 (0.79)\\
 84.420  & $^{2}\Pi_{3/2}$ $J$=7/2$^+$-5/2$^-$ & 2.38 (0.91)\\
163.396  & $^{2}\Pi_{1/2}$ $J$=3/2$^-$-1/2$^+$ & 1.34 (0.17)\\
163.015  & $^{2}\Pi_{1/2}$ $J$=3/2$^+$-1/2$^-$ & 2.74 (0.16)\\
 79.179  & $^{2}\Pi_{1/2}$-$^{2}\Pi_{3/2}$ $J$=1/2$^+$-3/2$^-$ & 9.17 (1.30)\\
 79.115  & $^{2}\Pi_{1/2}$-$^{2}\Pi_{3/2}$ $J$=1/2$^-$-3/2$^+$ & 9.65 (1.54)\\
 65.279  & $^{2}\Pi_{3/2}$ $J$=9/2$^+$-7/2$^-$ & 1.02 (0.49)\\
 65.131  & $^{2}\Pi_{3/2}$ $J$=9/2$^-$-7/2$^+$ & 1.32 (0.54)\\\hline\hline
\end{tabular}
\tablefoottext{a}{Near the OH emission peak:
$\alpha_{2000}$$\simeq$5$^h$35$^m$21.9$^s$, $\delta_{2000}$$\simeq$$-$5$^o$25$'$06.7$''$}
\label{table:intens}
\end{table}

\clearpage

\vspace{10cm}

\section{Online figures}
\label{sec:appenixB}

\begin{figure}[ht]
\begin{center}
\resizebox{\hsize}{!}{\includegraphics[angle=-90]{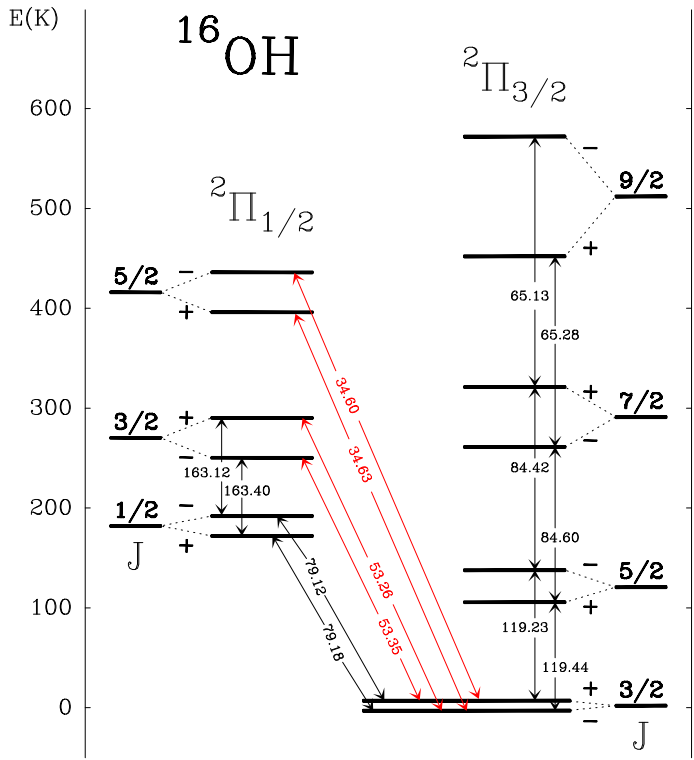}}
\end{center}
\caption{Rotational energy diagram of $^{16}$OH showing the lines detected by \textit{Herschel}/PACS (in microns). 
Detected transitions are shown in black while other pumping transitions discussed in the text
are shown in red. 
OH $^{2}\Pi_{3/2}$ and $^{2}\Pi_{1/2}$ rotational ladders are produced by the spin-orbit interaction, while the  
$\Lambda$-doubling splitting of each rotational level is produced by the nuclei rotation and
the unpaired electron motion. 
The $\Lambda$-doubling splitting
has been enhanced for clarity. Hyperfine structure is not shown.}
\label{fig_levels}
\end{figure}

\begin{figure}[bh]
\begin{center}
\includegraphics[width=8cm, angle=0]{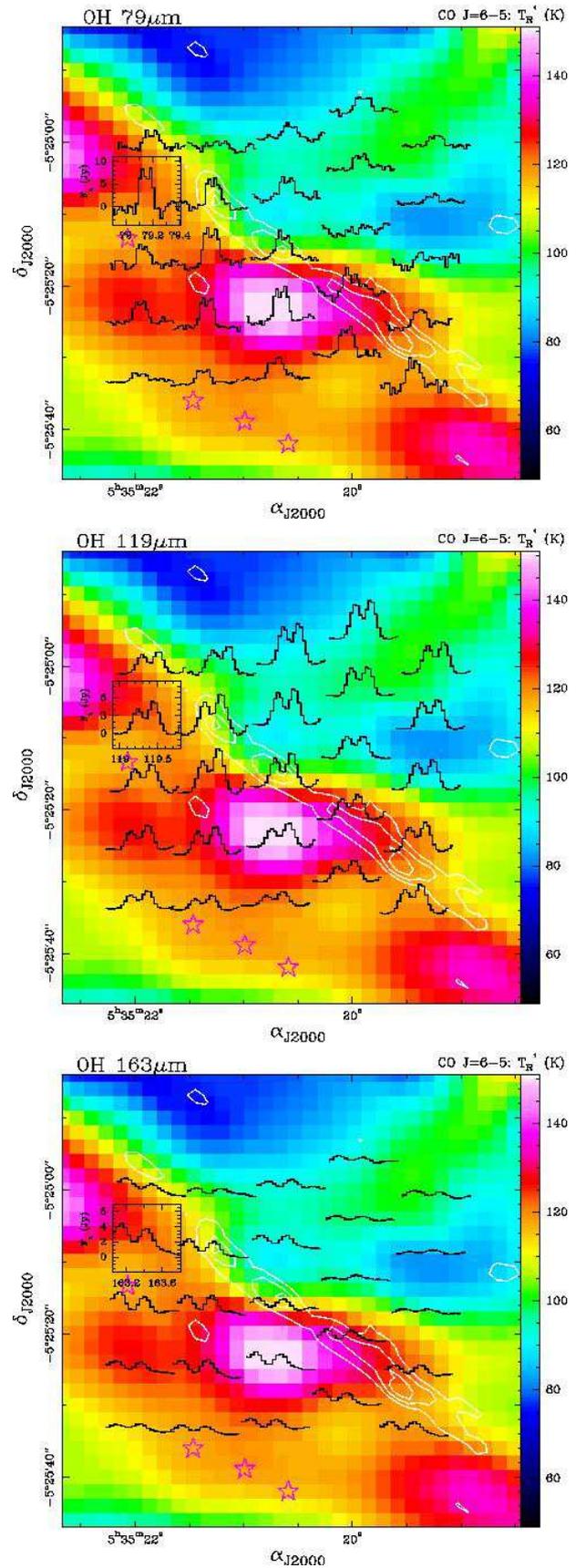}
\end{center}
\caption{Same as Figure~\ref{fig:OH_maps} for the 
OH $^{2}\Pi_{1/2}$-$^{2}\Pi_{3/2}$ $J$=1/2$\rightarrow$3/2 cross-ladder
lines at $\sim$79\,$\mu$m, the OH $^{2}\Pi_{3/2}$ $J$=5/2$\rightarrow$3/2
ground-state lines at $\sim$119\,$\mu$m, and the 
OH $^{2}\Pi_{1/2}$ $J$=3/2$\rightarrow$1/2
 excited lines at $\sim$163\,$\mu$m. Note the decrease of OH line intensity
with distance from the ionization front.
}
\label{fig:OH_maps2}
\end{figure}

\end{appendix}

\end{document}